\documentclass[12pt]{iopart}

\usepackage{graphicx} 
\usepackage[backend=biber,natbib=true, citestyle=numeric-comp,sorting=none,defernumbers=true,bibencoding=utf8,style=numeric]{biblatex}
\usepackage{color}
\usepackage[useregional]{datetime2}
\begin{document}

\title[ ]{Enhanced photocatalytic efficiency of layered CdS/CdSe heterostructures: Insights from first principles electronic structure calculations }
\author{Sulakhshana Shenoy \& Kartick Tarafder*}
\address{Department of Physics, National Institute of Technology, Srinivasnagar, Surathkal, Mangalore Karnataka-575025, India}
\ead{karticktarafder@gmail.com}
\vspace{10pt}
\begin{indented}
%\item[]March 2019
\item[]\today
\end{indented}
\begin{abstract}
Metal sulfides are emerging as an important class of materials for  photocatalytic applications, because of their high photo responsive nature in the wide visible light range. CdS in this class of materials, have a direct band gap of 2.4 eV, have gained special attention due to the relative position of its conduction band minimum, which is very close to the energies of the reduced protons. However, the photogenerated holes in the valence band of CdS are prone to oxidation and destroy its structure during photocatalysis. Thus constructing a CdS based heterostructure would be an effective strategy for improving the photocatalytic performance. In this work we have done a detail theoretical investigation based on  hybrid density functional theory calculation to get insight into the energy band structure, mobility and charge transfer across the CdS/CdSe heterojunction. The results indicate that CdS/CdSe forms type-II heterostructure that has several advantages in improving the photocatalytic efficiency under visible light irradiation. 
\end{abstract}

\section{Introduction}

The field of processing semiconductor-based photocatalysts has expanded rapidly in the last four decades. Notable developments have already been achieved especially in the area of organic pollutants decomposition and water splitting [1-7]. Most of the significant developments in this area are mainly based on titanium dioxide (TiO$_2$), which exhibits strong redox potential and excellent photocatalytic activity. However, a wide band gap of 3.2 eV in this material hinders its applicability as it can only be driven through UV light radiation which is only 3\% of solar spectrum [5,8-14]. Therefore designing and investigation of efficient visible light responsive materials for photocatalytic activity  has attracted researcher with a considerable attention from the viewpoint of cost, availability and environmental friendliness [15-20]. Few good visible light responsive semiconductor photocatalysts have recently been reported [15,16], among them cadmium sulfide (CdS) turns out to be very effective because of its relatively low band gap of 2.4 eV.  CdS belongs to II-VI group semiconductors, shows broad absorbance in the visible light region.  Also its relatively negative conduction band position with respect to water reduction potential make this material suitable for water spitting [21-23].  Efficient generation of charge carriers, charge separation and migration to the catalytic surface are the determining factors that affects the photocatalytic performance of a given semiconductor [24].   In photocatalytic reaction such as water splitting or photo-degradation of organic pollutants, the redox reactions on the semiconductor’s surface must be thermodynamically favorable. A photocatalytic reaction can only take place when the band edges of semiconductor are  appropriately placed relative to the redox reaction potentials. In particular, the position of the valence band maximum (VBM) and conduction band minimum (CBM) must lie lower and higher in energy relative the oxidation and reduction potentials respectively. In case of CdS, the photogenerated electrons in the conduction band have a good reducing ability for hydrogen evolution. However the photogenerated holes in the valence band are prone to self-oxidation, results to severe photo-corrosion in the system that  significantly reduces its photocatalytic efficiency. To overcome this drawback,  suitable band engineering is very essential which in turn help to separate the electron-hole pairs generated under visible light irradiation, decrease the rate of recombination and photo-corrosion.  Heterostructure formation  of CdS with some other efficient photocatalyst could be a suitable strategy for  further improvement of the photocatalytic efficiency and its durability. Cadmium selenide (CdSe) is one among the many other narrow band gap semiconductors having a direct band gap of 1.7 eV, capable of absorbing more of visible light spectrum, would be a good choice for making the heterostructure [25,26]. Experimental studies shows that combination of CdS/CdSe heterostructure could be a suitable candidate for  water splitting and photovoltaic applications, owing to its internal  charge separation at the interface [26,27]. Although an extensive experimental investigation have been carried out for  CdS/CdSe heterostructures towards its photocatalytic applications,  proper theoretical knowledge such as the detail electronic structure,   accurate position and character of the band edges are still lacking, which are very essential for a better understanding of catalytic mechanism such as light induced charge separation, electron and hole pair recombination etc. In this work, we have used hybrid density functional theory calculations to obtain an accurate electronic band structure of  CdS(110) and CdSe (110) surfaces as well as CdS/CdSe heterojunction and subsequently investigated  the possible improvement of photocatalytic activity with the help of their proper band alignment with respect to the reduction and oxidation potential.

\section{Computational Method} 
We employed density functional theory calculation using the projector augment wave (PAW) method implemented in  Vienna ab initio Simulation Package (VASP)  to determine the accurate geometric and electronic structure of the CdS, CdSe surfaces and CdS/CdSe heterostructure [28-30].  
For the exchange-correlation functional, we have used generalized gradient approximation (GGA) with  PBE parametrization. A very high wave function cutoff energy of 500 eV was used in each calculation to obtain accurate result.  The structure relaxation was performed in each case setting the convergence criteria $ 1 \times 10^{-5}$ eV for energy  and 0.01 eV\AA$^{-1}$  for force. Finally, we performed Hybrid-DFT calculation using Heyd-Scuseria-Ernzerhof (HSE06) hybrid functional along with GGA-PBE to extract accurate electronic structures of surfaces and heterostructure [31]. In all our calculations, the mixing parameters $\alpha$ in HSE06 are set as 0.25, which provides the calculated band gaps of systems are  close to their respective experimental values.

\section{Results and discussion}

Before exploring the properties of CdS/CdSe heterostructure, we first investigated the crystal structure of their bulk counterparts. Both CdS and CdSe belong to cubic sphalerite structure with {\sl F-43m} space group.   The optimized lattice parameters a = b = c = 5.82 \AA \ for CdS and  a =b = c = 6.07 \AA \ for CdSe are in good agreement with the experimental values [32]. Next we modeled CdS (110) and CdSe (110) surface, using optimized bulk geometry.  A vacuum region of 10\ \AA \ thickness perpendicular to the surface is used in the unit cell, to avoid the interactions between  neighboring slabs. The unit cell of an eight layer CdS (110) and CdSe (110) slabs are illustrated in Fig.1a and 1b. Finally the heterojunction was modeled by placing CdSe (110) slab over CdS (110) as shown in Fig.1c with a lattice mismatch less than 2.0\%.  We further relaxed the modeled surfaces and heterostructures to get the appropriate electronic structure information.  

\begin{figure}[!ht]
	\centering
	\includegraphics[width=0.7\linewidth]{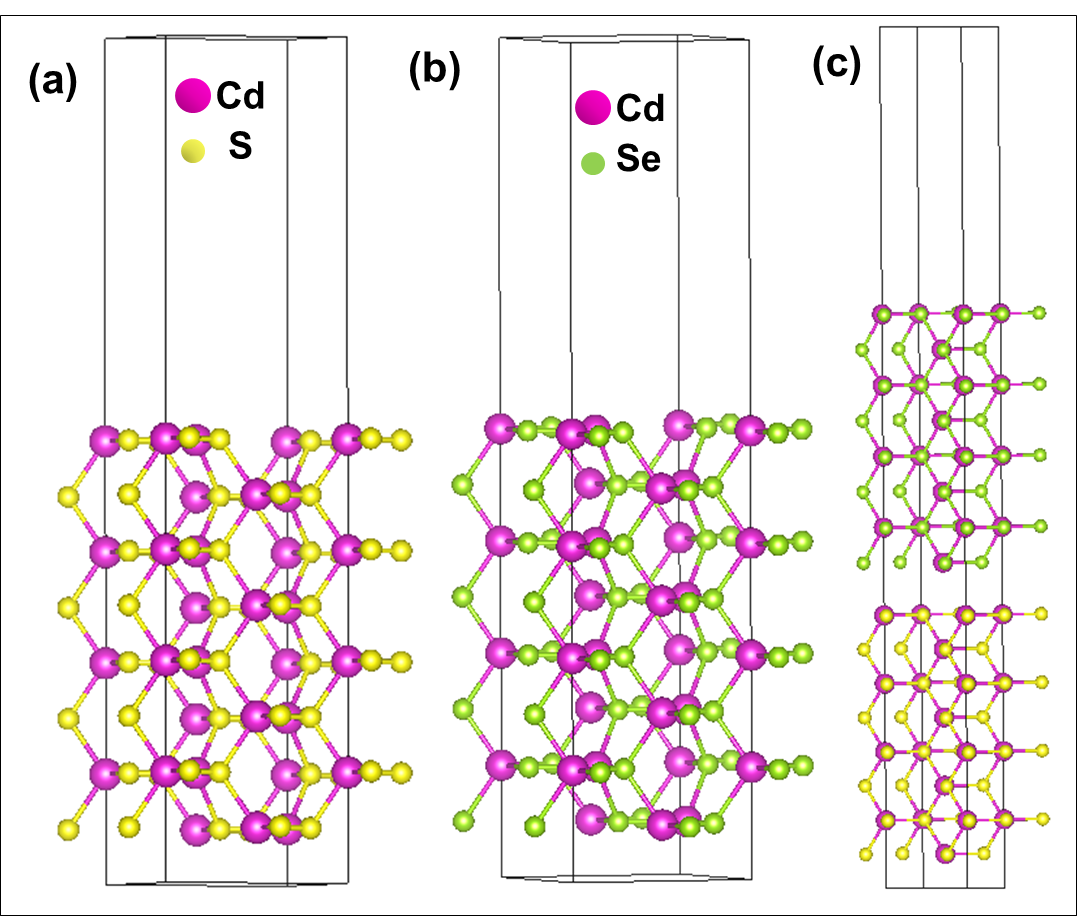}\quad
	\caption{(colour online)Optimized geometric structures of (a)CdS(110) surface, (b)CdSe(110) surface, and (c) CdS/CdSe heterojunction. Here pink, yellow and green balls are representing Cd, S and Se atom respectively}
	\label{fig1}
\end{figure}

The bulk CdS and CdSe are direct band gap semiconductors in which both the VBM and CBM are located at $\Gamma$ point. The calculated band gaps are 2.37 eV and 1.8 eV respectively, which are consistent with the values reported in the liturature [33]. In case of  CdS (110) and CdSe (110) surfaces, we found both the surfaces  have direct band gap of 2.39 eV and 1.8 eV respectively at the $\Gamma$ point. Our calculation shows that the CdS/CdSe heterostructure is also a direct gap system with band gap 1.83 eV at $\Gamma$ point.  Band structures of CdS(110),  CdSe(110) and the CdS/CdSe heterostructure  are shown in Fig.2a-c.

\begin{figure}[!ht]
	\centering
	\includegraphics[width=1.0\linewidth]{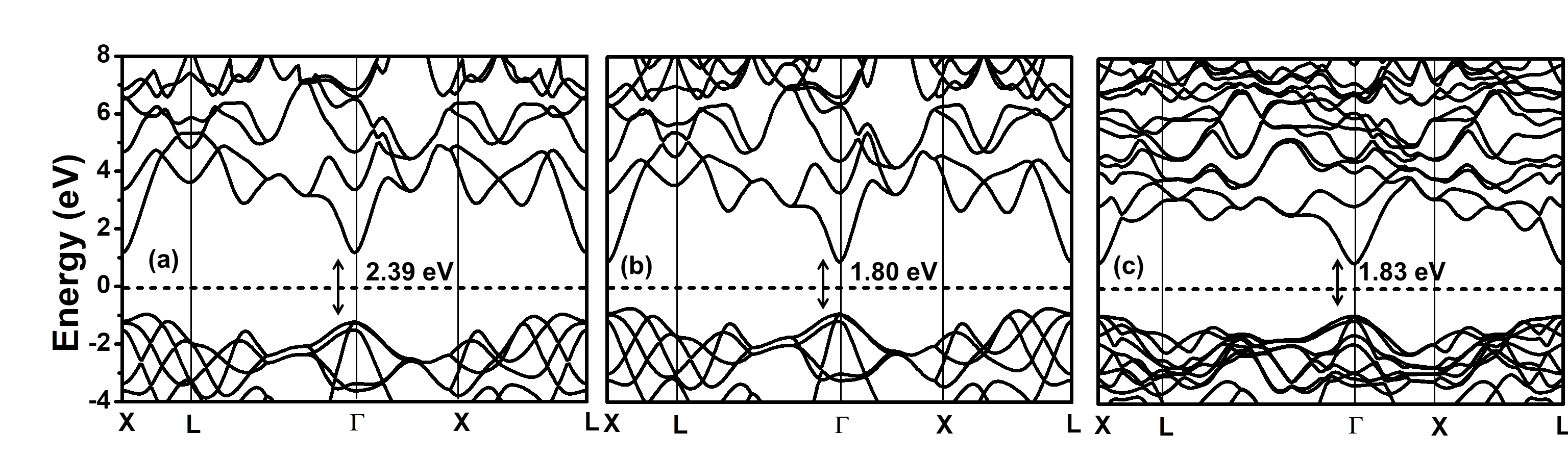}\quad
	\caption{(colour online) Calculated band structures of (a) CdS(110) (b)CdSe(110) surface and (c) CdS/CdSe heterostructure  using  hybrid-DFT calculation .}
	\label{fig2}
\end{figure}

To understand the orbital character of band edges, we have plotted the atom projected DOS for each system shown in Fig.3a-c.  From the atom projected DOS of CdS(110), it is evident that the VBM  consists of  S-3p states, whereas S-3s states are only contributed to the CBM. In case of CdSe(110) surface, Se 4p is contributing to the VBM, whereas Se-4s/4p states are contributing to  the CBM. In CdS/CdSe heterojunction, VBM is dominated by Se 4p states whereas CBM is of S-3s character. This confirms that it is a type-II semiconductor heterojunction.

 Next we investigated the catalytic activity for this materials, in which   relative band position of the heterostructure with respect to the reduction and oxidation potential is very essential. 
\begin{figure}[!ht]
	\centering
	\includegraphics[width=1.0\linewidth]{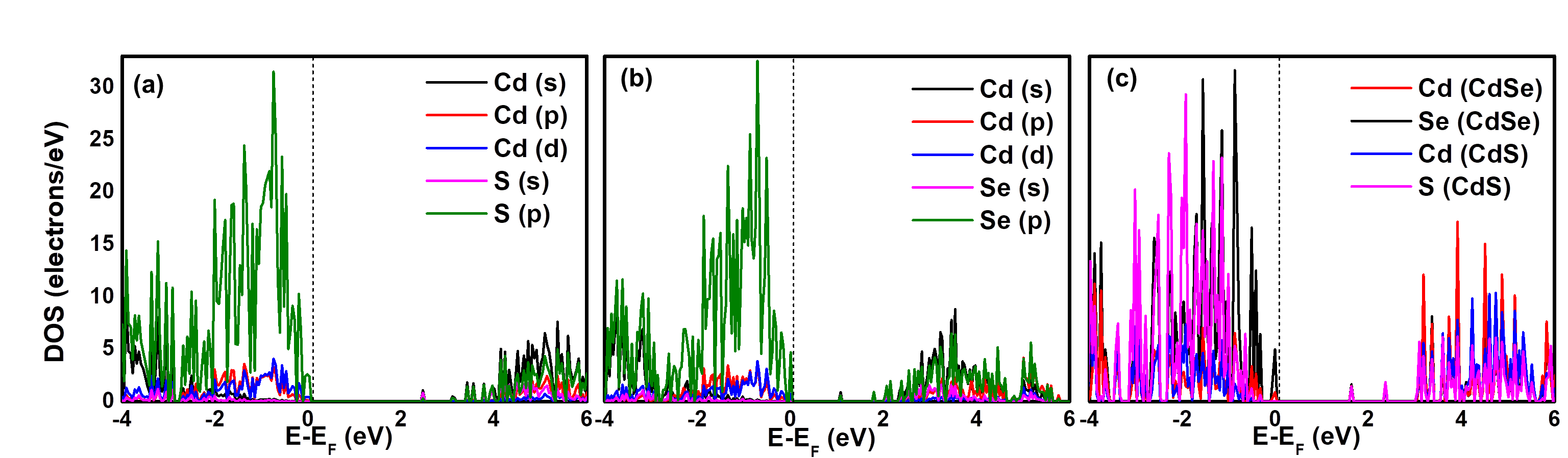}\quad
	\caption{(colour online) Atom projected density of state plots for (a)CdS(110) surface, (b) CdSe(110) surface and (c)CdS/CdSe heterostructure. }
	\label{fig3}
\end{figure}
The work function ($\Phi$) of material, in this case is a crucial parameter, commonly used as an intrinsic reference for band alignment [34].  This is the minimum energy required for  an electron to reach vacuum level from  Fermi level ($E_F$). One can estimate material specific work function from  the following equation:
\begin{equation}
\Phi=E_{vac}-E_F
\end{equation}
where $E_F$ is the Fermi energy. $E_{vac}$ is the energy of stationary electron in the vacuum nearby the surface. This can be estimated from the average electrostatic potential plot for a surface calculation, considering   sufficient amount of vacuum in the unit cell. Calculated electrostatic potential for CdS(110), CdSe(110) and  CdS/CdSe heterojunction surfaces  are plotted  in Fig.4a-c. On the basis of equation(1), our calculated work functions for CdS(110), CdSe(110) and CdS/CdSe heterojunction are 4.99 eV, 4.60 eV and 4.72 eV respectively. 
\begin{figure}[!ht]
	\centering
	\includegraphics[width=1.0\linewidth]{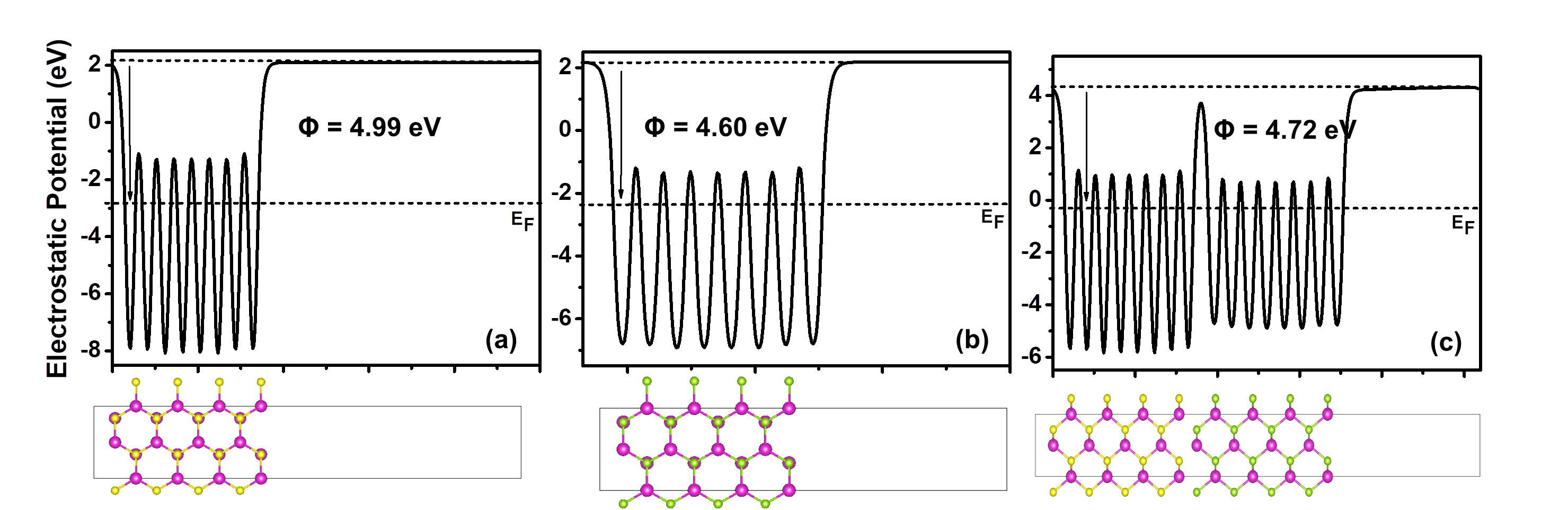}\quad
	\caption{(colour online)Average electrostatic potentials plots for (a) CdS(110) surface, (b) CdSe(110) surface and (c) CdS/CdSe heterojunction. Calculated work functions for CdS(110), CdSe(110) and CdS/CdSe are 4.99 eV, 4.60 eV and 4.72 eV respectively}
	\label{Optimized structure and PDOS for FG-Piperidine}
\end{figure}
The redox potential of a semiconductor is assessed by the positions of its valence and conduction band edges with respect to the normal hydrogen electrode (NHE) potential.  Generally, a positive valence band maximum of a semiconductor with respect to the NHE potential indicates that the photogenerated holes have stronger oxidation ability, while the negative conduction band minimum with respect to the NHE potential shows a strong reducing ability of the photogenerated electrons. Therefore accurate band edge position with respect to NHE will provide detail information about the photocatalytic efficiency of a semiconductor.  We have obtained the  edge positions of VB and CB for all three systems considering the Mulliken electronegativity rules [34], where the energetic position of the CBM and VBM is determined through following equations:
\begin{eqnarray}
E_{VB}=\chi-E_e+\frac{1}{2}E_g\\
E_{CB}=E_{VB}-E_g
\end{eqnarray}
Here $E_g$ represents the band gap of the system, $E_e$ is the energy of free electrons in the hydrogen scale (4.5 eV),  $E_{VB}$ and $E_{CB}$ are the VB and CB edge potentials, respectively [34].  $\chi$ in equation (2) is the Mulliken electronegativity of the semiconductor which  can be obtained by taking the geometric mean of the Mulliken electronegativity of constituent atoms in the semiconductor. The $\chi$ of an atom is the arithmetic mean of its electron affinity and the first ionization energy. The calculated values of $\chi$ for CdS and CdSe are 5.18 and 5.14 eV respectively. Using our calculated band gap values for CdS(110) and CdSe(110) along with equations (2) and (3), we have obtained the band edge positions of CB and VB  of CdS with respect to NHE as -0.51 V and 1.87 V  respectively and the band edge positions of CB and VB of CdSe are -0.26 V and 1.54 V respectively.  The results are illustrated in Fig.5.
\begin{figure}[!ht]
	\centering
	\includegraphics[width=0.80\linewidth]{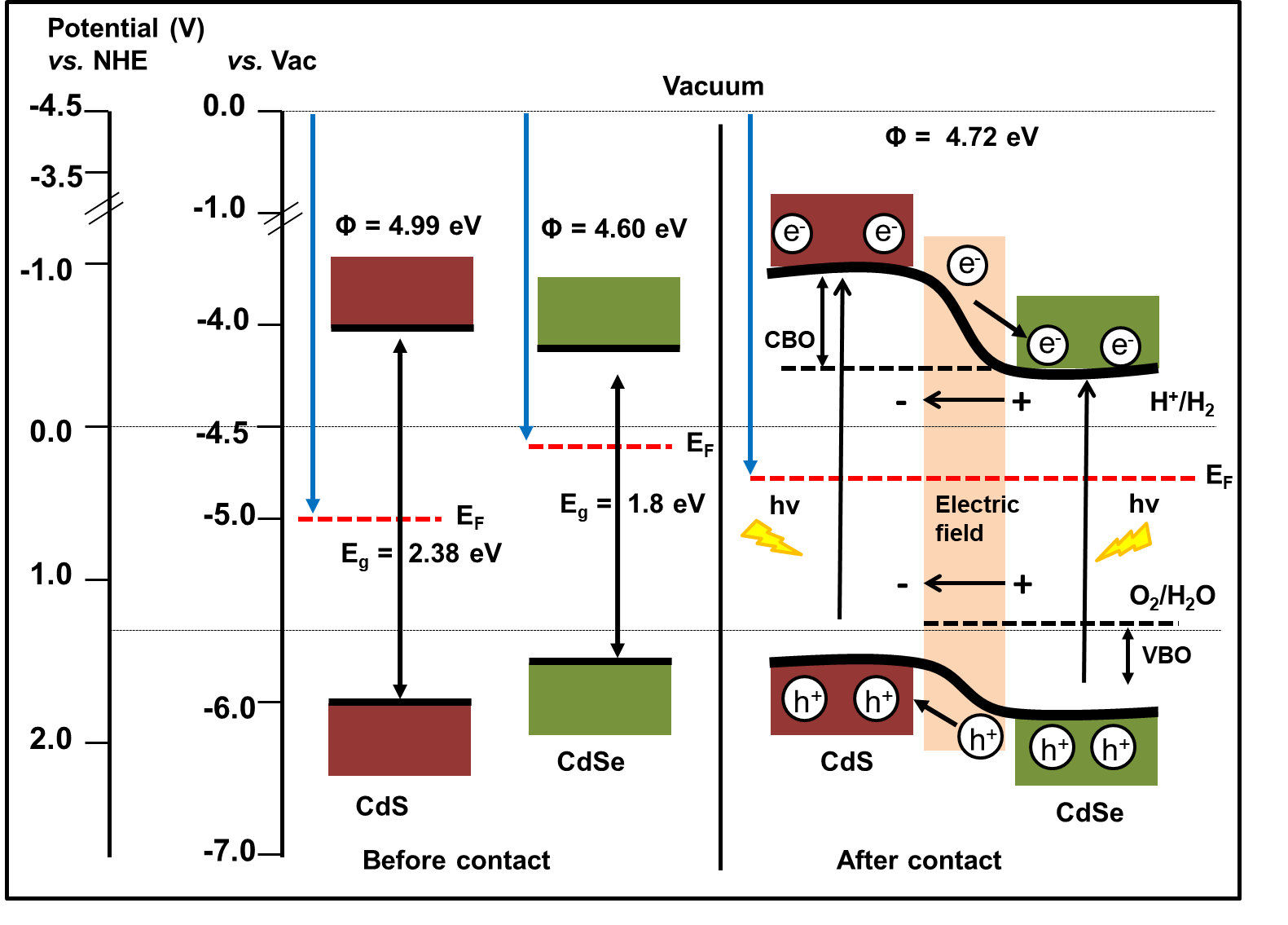}\quad
	\caption{(colour online) Diagram of band edge positions of CdS(110) and CdSe(110) surfaces  before and after the formation of heterojunction, with respect to the vacuum level}
	\label{fig5}
\end{figure}
 
In the case of CdS/CdSe heterojunction, the relative positions of CB and VB of both the semiconductors are expected to be change significantly because of the change of Fermi energy.   Since the work function of CdS (4.99 eV) is higher than CdSe (4.60 eV), electrons will flow from CdSe to CdS while forming the heterojunction, until the Fermi level of both the semiconductors are aligned.  Our calculation shows that the work function of the CdS/CdSe heterojunction is 4.72 eV, which is lying in between that of their individual semiconductors.  Therefore, at equilibrium the CB and VB of CdS  shift upwards by 0.49 eV, while CB and VB of CdSe shifted downwards by 0.1 eV due to the change of  Fermi level in the heterojunction and created a large band offsets both in valence and conduction bands. An internal charge separation is expected due to this band offsets. 

To confirm the charge separation at the interface of CdS/CdSe heterojunction, we have calculated the  charge density difference using the following equation
\begin{equation}
\Delta\rho=\rho_{CdS/CdSe}-\rho_{CdS}(110)-\rho_{CdSe}(110)
\end{equation}
               
here $\rho_{CdS/CdSe}$, $\rho_{CdS}(110)$ and $\rho_{CdSe}(110)$ represent the charge densities of CdS/CdSe heterostructure, the CdS(110) and the CdSe(110) surfaces respectively. The result shown in Fig.6(below), clearly indicates that the charge density is redistributed by the formation of hole-rich and electron-rich regions  near  the semiconductor interface.
\begin{figure}[!ht]
	\centering
	\includegraphics[width=0.80\linewidth]{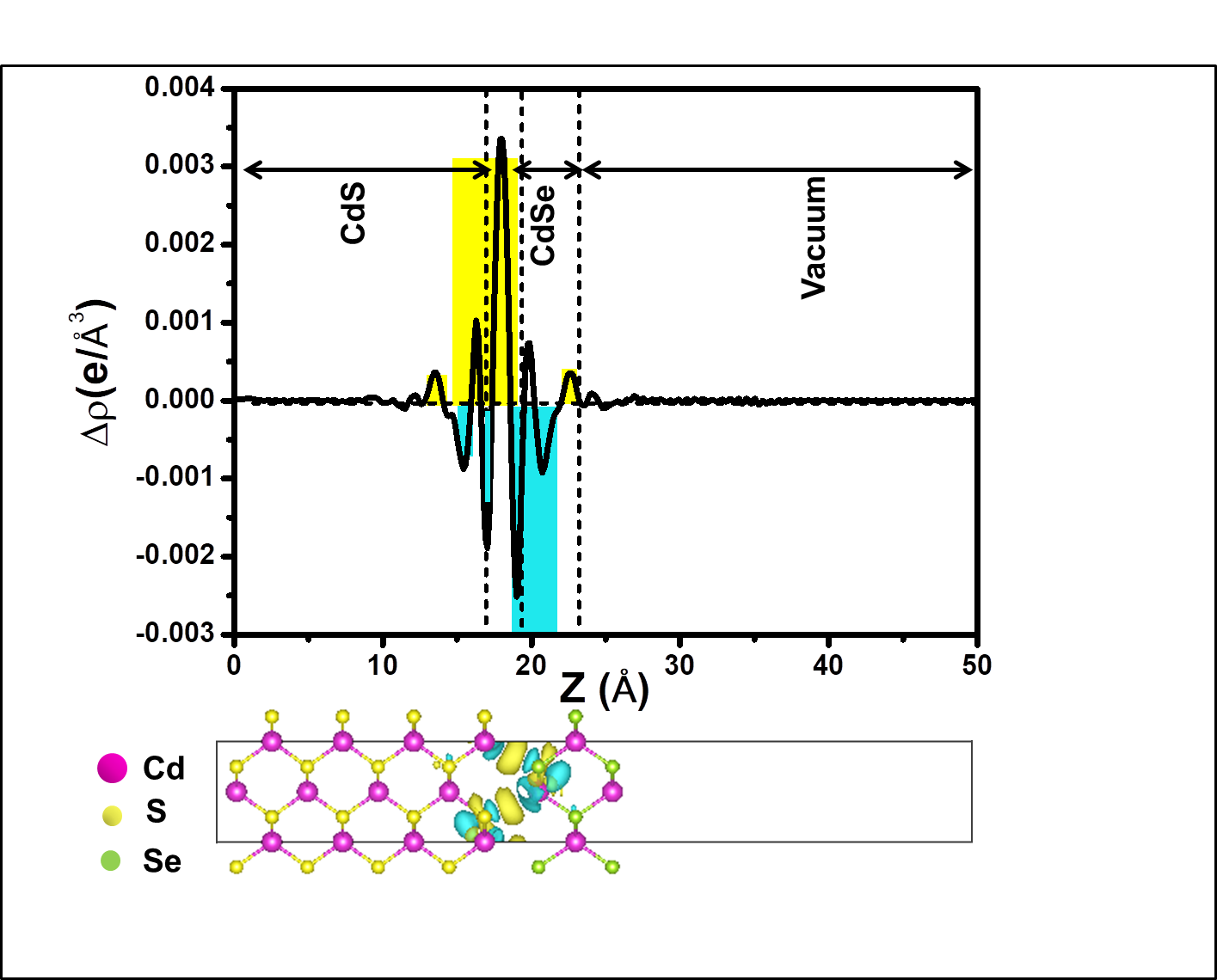}\quad
	\caption{(colour online)(above)  Planar-averaged electron density difference $\Delta\rho(z)$ for CdS/CdSe. The cyan and yellow region indicates electron depletion and electron accumulation, respectively. (below) Charge density difference plot for CdS/CdSe heterostructure}
	\label{fig6}
\end{figure}
 The accumulation of electrons in CdS is indicated by yellow isosurface, and electron depletion in the CdSe side is shown as cyan isosurface in Fig. 6. Note that the charge redistribution is confined in a small region close to the CdS/CdSe interface. There is no significant change is visible in the region farther away from the interface. This is because of the strong build up potential due to the separated charges, which acts against the diffusion process of the separated electron/holes. The planar-averaged electron density difference along the z-direction for the heterojunction is plotted in the Fig.6(above). indicates that the edge of CdS in the junction is populated with negative charge carrier, while CdSe edge is composed of positive charge carrier. This signifies that the charge separation leads to the formation of built-in electric field at the interface in the direction from CdSe to CdS during formation of CdS/CdSe heterojunciton which will help to separate the photo-generated charge carriers spatially across the heterojunction in opposite direction, thereby enhancing the possibility of higher photocatalytic performance.  
  
The calculated values of valence band offset (VBO) and conduction band offset (CBO) for CdS/CdSe heterojunction are 0.06 and 0.64 eV respectively. Under visible light irradiation, when the electron transition takes place from the VB to the CB of the heterojunction, owing to the existence of CBO, the electrons will start moving from CdS to CdSe. On the other hand, the VBO induces simultaneous movement of holes created in the VB from CdSe to CdS. As a result, negative charges get accumulated over CdSe and conversely positive charges on CdS. Thus, photoexcited electrons and holes are effectively separated and will participate in the photocatalytic reaction.  The photoexcited electrons of the CdS/CdSe heterojunction accumulated in the CB which is mainly contributed from CdSe. Since the CB edge potential is -0.36 V with respect to NHE, which is  more negative than that of H$^+$/H$_2$O (0 V),  has ability to reduce  the H$^+$ to H$_2$; while the photogenerated  holes accumulated at VB which is from CdS portion of the heterojunction  has edge potential of 1.38 V with respect to NHE, is more positive than that of O$_2$/H$_2$O (1.23 V), has very good oxidation ability to produce super-oxide anion radicals and degrade the organic pollutants. Therefore, CdS/CdSe becomes a type II heterojunction in which band edge positions are favorable for   simultaneous oxidation and reduction reactions.
\par
The mobility of charge carriers in a semiconductor photocatalyst has a greater influence on  its photocatalytic activity. Higher the mobility better is the performance of the photocatalyst. The mobility of charge carrier can be estimated from the band effective masses of a semiconductor as the  mobility of the charge carriers is inversely proportional to the effective mass. Therefore a lower value of effective mass promotes efficient migration of charge carriers and suppresses their recombination rate. We have estimated the effective masses of electrons and holes by fitting the parabolic portions to the CBM and VBM using the following equation:
\begin{equation}
m_x^*=\frac{\hbar}{2a}
\end{equation}

where $m_x^*(x=e,\  h)$ represents effective mass of  x-type charge carrier and $a$ is the coefficient of the second order term in a quadratic fit of E(K). A large difference in $m_e^*$ and   $m_h^*$ ie. larger value of $m_h^*$/$m_e^*$   in a semiconductor will  suppress the charge carrier recombination [35]. 
\begin{table}[!ht]
	\caption{\label{table1}Effective mass of charge carriers in CdSe(110),\\ CdSe(110) surface and CdS/CdSe heterostructure }
	\begin{indented}
		\item[]\begin{tabular}{@{}llll}
			\br
			\textbf {System} & \centering{$m^*_e$} & $m^*_h$ & \ $m^*_e$/$m^*_e$\\
			& \centering{(in e-mass)}&  \centering{(in e-mass)}            & \\ 
			\mr
			CdS(110)                 & 0.272   & 1.183 & 4.349 \\
			CdSe(110)                & 0.223   & 1.060 & 4.75 \\
			CdS/CdSe heterostructure & 0.210   & 1.230 & 5.857 \\ 
			\br
		\end{tabular}
	\end{indented}
\end{table}
Calculated effective masses are tabulated in Table 1.  Interestingly, our calculation shows that  value of $m_h^*$/$m_e^*$   is greater in heterostructure than that of individual semiconductor counterpart, indicating a larger lifetime of photogenerated charge carrier in  the  heterojunction that will further improve the  photocatalytic activity.

\section{Conclusion}

In summary, the electronic structures of the CdS(110), CdSe(110) and CdS/CdSe heterojunction were investigated on the basis of hybrid DFT method. Calculated band structure and atom projected DOS confirm that the CdS/CdSe forms a type-II heterostructure. The proper band edge positions and their alignment with respect to the NHE were determined from the calculated band gap and work function for each surface.  The band edge positions of both the semiconductors are changed with the Fermi level while formation of heterostructure and a large band offset both in valence and conduction band  was observed. Therefore, under the influence of visible light irradiation, the photogenerated electrons and holes could move from CB of CdS to that of CdSe and from VB of CdSe to that of CdS respectively. The existence of an internal  electric field near the interface due to large band offsets will facilitate the charge separation across the CdS/CdSe interface, which in turn reduce the recombination of electron-hole pairs. The charge carrier mobility is also improved in the heterostructure and the band alignment of the system is such that both photo-reduction and photo-oxidation processes associated with water splitting are energetically feasible. The results indicate that the CdS/CdSe heterojunction may have wide application in photocatalysis for pollutant degradation and water splitting.

Finally, our present work not only discloses a promising candidate for photocatalysis, but also provides a new pathway for investigating  novel and next generation heterojunctions for their applications in photocatalysis.

\section{Acknowledgement}
KT would like to acknowledge NITK-high performance computing facility and also would like thank DST-SERB(project no. SB/FTP/PS-032/2014 ) for the financial support.

\Bibliography{99}

\item Serpone N 2000 Photocatalysis {\it Kirk-Othmer Encyclopedia of Chemical Technology (American Cancer Society)}\\
\item Linsebigler A L, Lu Guangquan and Yates J T 1995 {\it Chem. Rev. } {\bf 95 } 735-58\\
\item Hoffmann M R, Martin S T, Choi Wonyong and Bahnemann D W 1995 {\it  Chem. Rev.} {\bf 95} 69-96\\
\item Asahi R, Morikawa T, Ohwaki T, Aoki K and Taga Y 2001 {\it Science} {\bf 293} 269-71\\
\item Schneider J, Matsuoka M, Takeuchi M, Zhang J, Horiuchi Y, Anpo M and Bahnemann D W 2014 {\it Chem. Rev.} {\bf 114} 9919-86\\
\item Liu J, Wang H and Antonietti M 2016 {\it Chem. Soc. Rev.} {\bf 45} 2308-26\\
\item Spasiano D, Marotta R, Malato S, Fernandez-Ibañez P and Di Somma I 2015 {\it Appl. Catal. B Environ.} {\bf 170-171} 90-123\\
\item Ola O and Maroto-Valer M M 2015 {\it J. Photochem. Photobiol. C Photochem. Rev.} {\bf 24} 16-42\\
\item He Y, Sutton N B, Rijnaarts H H H and Langenhoff A A M 2016 {\it Appl. Catal. B Environ.} {\bf 182} 132-41\\
\item Marinho B A, Djellabi R, Cristóvão R O, Loureiro J M, Boaventura R A R, Dias M M, Lopes J C B and Vilar V J P 2017 {\it Chem. Eng. J.} {\bf 318} 76-88\\
\item Farner Budarz J, Turolla A, Piasecki A F, Bottero J-Y, Antonelli M and Wiesner M R 2017 {\it Langmuir} {\bf 33} 2770-9\\
\item Pueyo N, Miguel N, Mosteo R, Ovelleiro J L and Ormad M P 2017 {\it J. Environ. Sci. Health Part A} {\bf 52} 182-8\\
\item Low J, Cheng B and Yu J 2017 {\it Appl. Surf. Sci.} {\bf 392} 658-86\\
\item Liu N, Chen X, Zhang J and Schwank J W 2014 {\it Catal. Today} {\bf 225} 34-51\\
\item Schultz D M and Yoon T P 2014 {\it Science} {\bf 343} 1239176\\
\item Rao M P, Sathishkumar P, Mangalaraja R V, Asiri A M, Sivashanmugam P and Anandan S 2018 {\it J. Environ. Chem. Eng.} {\bf 6} 2003-10\\
\item Low J, Yu J, Jaroniec M, Wageh S and Al‐Ghamdi A A 2017 {\it Adv. Mater.} {\bf 29} 1601694\\
\item Zhao H, Sun S, Jiang P and Xu Z J 2017 {\it Chem. Eng. J.} {\bf 315} 296-303\\
\item Zhao H, Sun S, Wu Y, Jiang P, Dong Y and Xu Z J 2017 {\it Carbon} {\bf 119} 56-61\\
\item Lin F, Shao Z, Li P, Chen Z, Liu X, Li M, Zhang B, Huang J, Zhu G and Dong B 2017 {\it RSC Adv.} {\bf 7} 15053-9\\
\item Zhou P, Le Z, Xie Y, Fang J and Xu J 2017 {\it J. Alloys Compd.} {\bf 692} 170-7\\
\item Bera R, Kundu S and Patra A 2015 { \it ACS Appl. Mater. Interfaces} {\bf 7} 13251-9\\
\item Garg P, Kumar S, Choudhuri I, Mahata A and Pathak B 2016 {\it  J. Phys. Chem. C} {\bf 120} 7052-60\\
\item Cheng L, Xiang Q, Liao Y and Zhang H 2018 {\it Energy Environ. Sci.} {\bf 11} 1362-91\\
\item Chauhan H, Kumar Y, Dana J, Satpati B, Ghosh H N and Deka S 2016 {\it Nanoscale} {\bf 8} 15802-12\\
\item Bera R, Dutta A, Kundu S, Polshettiwar V and Patra A 2018 {\it  J. Phys. Chem. C} {\bf 122} 12158-67\\
\item Bridewell V L, Alam R, Karwacki C J and Kamat P V 2015 {\it Chem. Mater.} {\bf 27} 5064-71\\
\item Kresse G and Furthmüller J 1996 {\it Comput. Mater. Sci.} {\bf 6} 15-50\\
\item Blöchl  null 1994 {\it Phys. Rev. B Condens. Matter} {\bf 50} 17953-79\\
\item Perdew J P and Wang Y 1992 {\it Phys. Rev. B} {\bf 45} 13244-9\\
\item Heyd J, Scuseria G E and Ernzerhof M 2003 {\it J. Chem. Phys.} {\bf 118} 8207-15\\
\item Wei S-H and Zhang S B 2000 {\it Phys. Rev. B} {\bf 62} 6944-7\\
\item Mo Y, Tian G and Tao J 2017 {\it Chem. Phys. Lett.} {\bf 682} 38-42\\
\item Liu J 2015 {\it  J. Phys. Chem. C} {\bf 119} 28417-23\\
\item Chen X, Shen S, Guo L and Mao S S 2010 {\it Chem. Rev.} {\bf 110} 6503-70\\

\endbib

\end{document}